\documentstyle{npbproc}

\newcommand{\be}{\begin{equation}}
\newcommand{\ee}{\end{equation}}
\newcommand{\bea}{\begin{eqnarray}}
\newcommand{\eea}{\end{eqnarray}}
\newcommand{\rar}{\rightarrow}

\begin{document}

\title{ MIRROR FERMIONS IN CHIRAL GAUGE THEORIES }

\author{ I. Montvay }
\address{ Deutsches Elektronen-Synchrotron DESY, \\
          D-2000 Hamburg 52, FRG }

\date{}

\runtitle{ Mirror fermions }
\runauthor{ I. Montvay }

\pubyear{1992}
\volume{25A}
\firstpage{1}
\lastpage{12}

\begin{abstract}
 Mirror fermions appear naturally in lattice formulations of the
 standard model.
 The phenomenological limits on their existence and discovery
 limits at future colliders are discussed.
 After an introduction of lattice actions for chiral Yukawa-models,
 a recent numerical simulation is presented.
 In particular, the emerging phase structure and features of the
 allowed region in renormalized couplings are discussed.
\end{abstract}

\maketitle

\section{ INTRODUCTION }
 In the presently known energy range the spectrum of the
 elementary particles is {\em ``chiral''} in the sense that no
 explicit fermion mass terms are allowed by the symmetry.
 Fermion masses, as well as all other masses, are entirely
 generated by spontaneous symmetry breaking, due to the nonzero
 vacuum expectation value $v_R \simeq 250$ GeV of the Higgs-boson
 field.
 One of the consequences is the V-A structure of weak interactions,
 resulting in the breaking of space-reflection (parity) symmetry.

 A natural question is whether ``chirality'' and the accompanying
 parity breaking is perhaps only a low-energy phenomenon, and at
 high energy the space-reflection symmetry is restored by the
 existence of opposite chirality {\em ``mirror fermions''}
 \cite{LEEYAN}.
 If the presently known (almost complete) three fermion families
 were duplicated at the electroweak energy scale, in the range
 100-1000 GeV, by three mirror fermion families with opposite
 chiralities and hence V+A couplings to the weak gauge vector
 bosons \cite{MIRFAM}, then the whole fermion spectrum would be
 ``vectorlike''.
 This would very much simplify the nonperturbative lattice formulation
 of the Standard Model \cite{CHFER,TSUKPR}.
 Of course, since no effects of the mirror fermions are experimentally
 observed up to now, first one has to ask whether the limits implied
 by the presently known experimental data allow their existence at all.

\section{ MIRROR FERMION PHENOMENOLOGY }
 The mirror partners of fermions have the same
 $\rm SU(3) \otimes SU(2)_L \otimes U(1)_Y$ quantum numbers but
 opposite chiralities.
 For instance, the right-handed chiral components of mirror leptons
 form a doublet with $Y=-1$ with respect to
 $\rm SU(2)_L \otimes U(1)_Y$.
 Such particles appear in several extensions of the minimal Standard
 Model, for instance, in grand unified theories with large groups
 as O(16) or SU(15) \cite{MAAROO}.
 In general mirror fermion models there might be some other quantum
 numbers which are different for fermions and mirror fermions, and the
 set of representations containing the mirror fermions may also be
 different.
 The lattice formulation of the Standard Model suggests a simple
 doubling of the fermion spectrum \cite{NIENIN}, resulting in three
 mirror pairs of fermion families \cite{MIRFAM}.

\subsection{ Present limits }
 The direct pair production of mirror fermions is not observed at LEP.
 This puts a lower limit on their masses of about 45 GeV.
 Heavier mirror fermions could be produced via their mixing to
 ordinary fermions.
 This implies some constraints on the mixing angles versus the masses.

 In order to discuss the mixing schemes, let us first consider the
 simplest case of a single fermion ($\psi$) mirror fermion ($\chi$)
 pair.
 The mass matrix on the $(\overline{\psi}_R,\overline{\psi}_L,
 \overline{\chi}_R,\overline{\chi}_L) \otimes
 (\psi_L,\psi_R,\chi_L,\chi_R)$ basis is
\be \label{eq01}
M = \left( \begin{array}{cccc}
\mu_\psi  &  0  &  \mu_R  &  0       \\
0  &  \mu_\psi  &  0  &  \mu_L       \\
\mu_L  &  0  &  \mu_\chi  &  0       \\
0  &  \mu_R  &  0  &  \mu_\chi
\end{array} \right) \ .
\ee
 Here $\mu_{(L,R)}$ are the fermion mirror fermion mixing mass
 parameters, and the diagonal elements are produced by spontaneous
 symmetry breaking:
\be \label{eq02}
 \mu_\psi=G_{R\psi}v_R \ , \hspace{2em} \mu_\chi=G_{R\chi}v_R \ ,
\ee
 with the renormalized Yukawa-couplings $G_{R\psi}$, $G_{R\chi}$.

 For $\mu_R \ne \mu_L$ the mass matrix $M$ in (\ref{eq01})  is not
 symmetric, hence one has to diagonalize $M^T M$ by
 $O^T_{(LR)} M^T M O_{(LR)}$, and
 $M M^T$ by $O^T_{(RL)} M M^T O_{(RL)}$, where
$$
O_{(LR)} =
$$
$$
           \left( \begin{array}{cccc}
\cos\alpha_L  &  0  &  \sin\alpha_L  &  0   \\
0  &  \cos\alpha_R  &  0  &  \sin\alpha_R   \\
-\sin\alpha_L  &  0  &  \cos\alpha_L  &  0  \\
0  &  -\sin\alpha_R  &  0  &  \cos\alpha_R
\end{array} \right) \  ,
$$
$$
O_{(RL)} =
$$
\be \label{eq03}
           \left( \begin{array}{cccc}
\cos\alpha_R  &  0  &  \sin\alpha_R  &  0   \\
0  &  \cos\alpha_L  &  0  &  \sin\alpha_L   \\
-\sin\alpha_R  &  0  &  \cos\alpha_R  &  0  \\
0  &  -\sin\alpha_L  &  0  &  \cos\alpha_L
\end{array} \right) \ .
\ee
 The rotation angles of the left-handed, respectively, right-handed
 components satisfy
$$
\tan(2\alpha_L) = \frac{2(\mu_\chi \mu_L + \mu_\psi \mu_R)}
{\mu_\chi^2 + \mu_R^2 - \mu_\psi^2 - \mu_L^2} \ ,
$$
\be \label{eq04}
\tan(2\alpha_R) = \frac{2(\mu_\chi \mu_R + \mu_\psi \mu_L)}
{\mu_\chi^2 + \mu_L^2 - \mu_\psi^2 - \mu_R^2} \ ,
\ee
 and the two (positive) mass-squared eigenvalues are given by
$$
\mu_{1,2}^2 = \half \left\{
\mu_\chi^2 + \mu_\psi^2 + \mu_L^2 + \mu_R^2
\right.
$$
$$
\mp \left[ (\mu_\chi^2 - \mu_\psi^2)^2 + (\mu_L^2 - \mu_R^2)^2
\right.
$$
\be \label{eq05}
\left.\left.
+ 2(\mu_\chi^2 + \mu_\psi^2) (\mu_L^2 + \mu_R^2)
+ 8\mu_\chi \mu_\psi \mu_L \mu_R \right]^\half \right\}  .
\ee
 The mass matrix itself is diagonalized by
$$
O^T_{(RL)} M O_{(LR)} = O^T_{(LR)} M^T O_{(RL)}
$$
\be \label{eq06}
= \left( \begin{array}{cccc}
\mu_{1}  &  0  &  0  &  0   \\
0  &  \mu_{1}  &  0  &  0   \\
0  &  0  &  \mu_{2}  &  0   \\
0  &  0  &  0  &  \mu_{2}
\end{array} \right) \ .
\ee
 This shows that for $\mu_\psi,\mu_L,\mu_R \ll \mu_\chi$ there
 is a light state with mass $\mu_{1}=O(\mu_\psi,\mu_L,\mu_R)$ and a
 heavy state with mass $\mu_{2}=O(\mu_\chi)$.
 In general, both the light and heavy states are mixtures of
 the original fermion and mirror fermion.
 According to (\ref{eq04}), for $\mu_L \ne \mu_R$ the
 fermion-mirror-fermion mixing angle in the left-handed sector is
 different from the one in the right-handed sector.

 In case of three mirror pairs of fermion families the diagonalization
 of the mass matrix is in principle similar but, of course, more
 complicated.
 A particular class of mixing schemes will be discussed in the next
 subsection.
 The mirror fermions can be produced through their mixing to ordinary
 fermions.
 The upper limits on the absolute value of mixing angles depend on the
 mixing scheme.

 Indirect limits on the existence of heavy mirror fermions can also be
 deduced from the absence of observed effects in 1-loop radiative
 corrections \cite{PESTAK,ALTBAR}, because of the non-decoupling of
 heavy fermions.
 The question of non-decoupling in higher loop orders is, however,
 open.
 In fact, one of the goals of nonperturbative lattice studies is to
 investigate this in the nonperturbative regime of couplings.

\subsection{ Mixing schemes }
 The strongest constraints on mixing angles between ordinary
 fermions and mirror fermions arise from the
 conservation of $e$-, $\mu$- and $\tau$- lepton numbers and from
 the absence of flavour changing neutral currents \cite{MAAROO}.
 In a particular scheme these constraints can be avoided \cite{MIRFAM}.
 In this {\em ``monogamous mixing''} scheme the structure of the
 mass matrix is such that there is a one-to-one correspondence between
 fermions and mirror fermions, due to the fact that the family
 structure of the mass matrix for mirror fermions is closely related
 to the one for ordinary fermions.

 Let us denote doublet indices by $A=1,2$, colour indices by
 $c=1,2,3$ in such a way that the leptons belong to the fourth value of
 colour $c=4$, and family indices by $K=1,2,3$.
 In general the entries of the mass matrix for three mirror pairs
 of fermion families are diagonal in isospin and colour, hence they
 have the form
$$
\mu_{(\psi,\chi);A_2c_2K_2,A_1c_1K_1} = \delta_{A_2A_1}
\delta_{c_2c_1} \mu^{(A_1c_1)}_{(\psi,\chi);K_2K_1}  \ ,
$$
$$
\mu_{L;A_2c_2K_2,A_1c_1K_1} = \delta_{A_2A_1}
\delta_{c_2c_1} \mu^{(c_1)}_{L;K_2K_1}  \ ,
$$
\be \label{eq07}
\mu_{R;A_2c_2K_2,A_1c_1K_1} = \delta_{A_2A_1}
\delta_{c_2c_1} \mu^{(A_1c_1)}_{R;K_2K_1}  \ .
\ee
 The diagonalization of the mass matrix can be achieved for given
 indices $A$ and $c$ by two $6 \otimes 6$ unitary matrices $F_L^{(Ac)}$
 and $F_R^{(Ac)}$ acting, respectively, on the L-handed and R-handed
 subspaces:
$$
F_L^{(Ac)\dagger}(M^\dagger M)_L F_L^{(Ac)} \ ,
$$
\be \label{eq08}
F_R^{(Ac)\dagger}(M^\dagger M)_R F_R^{(Ac)} \ .
\ee

 The main assumption of the ``monogamous'' mixing scheme is that
 in the family space $\mu_\psi,\mu_\chi,\mu_L,\mu_R$ are hermitian
 and simultaneously diagonalizable, that is
\be \label{eq09}
F_L^{(Ac)} = F_R^{(Ac)} =  \left(
\begin{array}{cc}
F^{(Ac)}  &  0  \\  0  &  F^{(Ac)}
\end{array} \right)  \ ,
\ee
 where the block matrix is in $(\psi,\chi)$-space.
 The Kobayashi-Maskawa matrix of quarks is given by
\be \label{eq10}
C \equiv F^{(2c)\dagger} F^{(1c)} \ ,
\ee
 independently for $c=1,2,3$.
 The corresponding matrix with $c=4$ and $A=1 \leftrightarrow 2$
 describes the mixing of neutrinos, if the Dirac-mass of the neutrinos
 is nonzero.
 (Majorana masses of the neutrinos are not considered here, but
 in principle, they can also be introduced.)

 A simple example for the ``monogamous'' mixing is the following:
$$
\mu^{(Ac)}_{\chi;K_2K_1} = \lambda^{(Ac)}_\chi
\mu^{(Ac)}_{\psi;K_2K_1} + \delta_{K_2K_1}\Delta^{(Ac)} \ ,
$$
$$
\mu^{(c)}_{L;K_2K_1} = \delta_{K_2K_1}\delta^{(c)}_L \ ,
$$
\be \label{eq11}
\mu^{(Ac)}_{R;K_2K_1} = \lambda^{(Ac)}_R
\mu^{(Ac)}_{\psi;K_2K_1} + \delta_{K_2K_1}\delta^{(Ac)}_R \ ,
\ee
 where $\lambda_\chi^{(Ac)},\; \Delta^{(Ac)},\; \delta_L^{(c)},\;
 \lambda_R^{(Ac)},\; \delta_R^{(Ac)}$ do not depend on the family
 index.

 The general case can be parametrized by the eigenvalues
 $\mu^{(AcK)}_\psi,\; \mu^{(AcK)}_\chi,\; \mu^{(AcK)}_R,\;
 \mu^{(cK)}_L$ and matrices $F^{(Ac)}$:
$$
\mu^{(Ac)}_{(\psi,\chi,R);K_2K_1} = \sum_K
F^{(Ac)}_{K_2K} \mu^{(AcK)}_{(\psi,\chi,R)} F^{(Ac)\dagger}_{KK_1} \ ,
$$
\be \label{eq12}
\mu^{(c)}_{L;K_2K_1} = \sum_K
F^{(Ac)}_{K_2K} \mu^{(cK)}_L F^{(Ac)\dagger}_{KK_1} \ .
\ee
 Here in the second line the left hand side has to be independent of
 the value of $A=1,2$.

 The full diagonalization of the mass matrix on the
 $(\psi_L,\psi_R,\chi_L,\chi_R)$ basis of all three family pairs is
 achieved by the $96 \otimes 96$ matrix
$$
{\cal O}^{(LR)}_{A^\prime c^\prime K^\prime,AcK}
= \delta_{A^\prime A}\delta_{c^\prime c}
F^{(Ac)}_{K^\prime K}
$$
$$
\cdot \left(
\begin{array}{cc}
 \cos\alpha^{(AcK)}_L  &  0  \\
 0  &  \cos\alpha^{(AcK)}_R  \\
-\sin\alpha^{(AcK)}_L  &  0  \\
 0  & -\sin\alpha^{(AcK)}_R
\end{array}  \right.
$$
\be \label{eq13}
\left.
\begin{array}{cc}
 \sin\alpha^{(AcK)}_L  &  0  \\
 0  &  \sin\alpha^{(AcK)}_R  \\
 \cos\alpha^{(AcK)}_L  &  0  \\
 0  &  \cos\alpha^{(AcK)}_R
\end{array}  \right) \ .
\ee
 $M^\dagger M$ is diagonalized by
\be \label{eq14}
{\cal O}^{(LR)\dagger} M^\dagger M {\cal O}^{(LR)} \ ,
\ee
 and $M M^\dagger$ by
\be \label{eq15}
{\cal O}^{(RL)\dagger} M M^\dagger {\cal O}^{(RL)} \ ,
\ee
 where ${\cal O}^{(RL)}$ is obtained from ${\cal O}^{(LR)}$ by
 $\alpha_L \leftrightarrow \alpha_R$.

 In case of $\mu_R=\mu_L$, which happens for instance in (\ref{eq11})
 if $\lambda_R=0$ and $\delta_R=\delta_L$, the left-handed and
 right-handed mixing angles are the same:
\be \label{eq16}
\alpha^{(AcK)} \equiv \alpha_L^{(AcK)} =
\alpha_R^{(AcK)} \ .
\ee
 In Ref.~\cite{MIRFAM} only this special case was considered.
 The importance of the left-right-asymmetric mixing was pointed out in
 Ref.~\cite{CSIFOD}, where the constraints arising from the measured
 values of anomalous magnetic moments were determined.
 It turned out that for the electron and muon the upper limit is
\be \label{eq17}
|\alpha_L \alpha_R| \le 0.0004 \ ,
\ee
 which is much stronger than the limits obtained from all other
 data \cite{LANLON}:
\be \label{eq18}
\alpha_L^2,\; \alpha_R^2 \le 0.02 \ .
\ee
 In case of the L-R asymmetric mixing the constraint (\ref{eq17})
 can be satisfied, if either the left- or right-handed mixing exactly
 vanishes (or is very small): $\alpha_L=0$ or $\alpha_R=0$.

\subsection{ Future colliders }
 The hypothetical mirror fermions can be discovered at the next
 generation of high energy colliders.
 At HERA the first family mirror fermions can be produced via mixing
 to ordinary fermions up to masses of about 200 GeV, if the mixing
 angles are close to their present upper limits \cite{CSIMON,HERA}.
 At SSC and LHC mirror lepton pair production can be observed up to
 masses of a few hundred GeV \cite{CSIKOR}.
 This has the advantage of being essentially independent of the small
 mixing.
 At a high energy $e^+e^-$ collider, e.~g. LEP-200 or NLC,
 mirror fermions can be pair produced and easily identified up to
 roughly half of the total energy, and also produced via mixing
 almost up to the total energy \cite{NLC}.

\section{ LATTICE SIMULATION OF YUKAWA MODELS }

\subsection{ Lattice actions }
 The lattice formulation of the electroweak Standard Model is
 difficult because of the doubler fermions \cite{NIENIN}.
 In fact, at present no completely satisfactory formulation is known
 \cite{TSUKPR}: if one insists on explicit chiral gauge invariance,
 then mirror fermion fields have to be introduced \cite{CHFER},
 otherwise one has to fix the gauge as in the ``Rome-approach''
 \cite{ROMA2}.

 The situation is different if the
 $\rm SU(3)_{colour} \otimes U(1)_{hypercharge}$
 gauge couplings are neglected.
 In this case, as a consequence of the pseudo-reality of SU(2)
 representations, mirror fermions can be transformed to normal fermions
 by charge conjugation.
 This allows the gauge invariant lattice formulation of $\rm SU(2)_L$
 symmetric models with an even number of fermion doublets.
 For simplicity, let us also neglect here the $\rm SU(2)_L$ gauge
 interaction, and consider a chiral Yukawa-model of two fermion
 doublets.
 A global $\rm SU(2)_L \otimes U(1)$ symmetric chiral Yukawa-model
 can be formulated by the lattice action
\be \label{eq19}
S = S_{scalar}+ S_{fermion} \ ,
\ee
 where the pure scalar part in terms of the $2 \otimes 2$ matrix
 Higgs-boson field $\varphi$ is
$$
S_{scalar} = \frac{1}{4}\sum_x \left\{
m_0^2 {\rm Tr\,}(\varphi^\dagger_x\varphi_x)
+ \lambda \left[ {\rm Tr\,}(\varphi^\dagger_x\varphi_x) \right]^2
\right.
$$
\be \label{eq20}
\left.
+  \sum_\mu
[ {\rm Tr\,}  (\varphi^\dagger_x \varphi_x)
- {\rm Tr\,}  (\varphi^\dagger_{x+\hat{\mu}}\varphi_x) ]
\right\} \ ,
\ee
 and the fermionic part with two doublet fields $\psi_{1,2}$ is
$$
S_{fermion} = \sum_x \left\{ \frac{\mu_0}{2} \left[
  (\psi^T_{2x}\epsilon C \psi_{1x}) - (\psi^T_{1x}\epsilon C \psi_{2x})
\right.\right.
$$
$$
\left.
+ (\overline{\psi}_{2x}\epsilon C \overline{\psi}^T_{1x})
- (\overline{\psi}_{1x}\epsilon C \overline{\psi}^T_{2x}) \right]
$$
$$
- \half \sum_\mu \left[
  (\overline{\psi}_{1 x+\hat{\mu}} \gamma_\mu \psi_{1x})
+ (\overline{\psi}_{2 x+\hat{\mu}} \gamma_\mu \psi_{2x})
\right.
$$
$$
- \frac{r}{2} \left(
  (\psi^T_{2x}\epsilon C \psi_{1x})
- (\psi^T_{2 x+\hat{\mu}}\epsilon C \psi_{1x})
\right.
$$
$$
- (\psi^T_{1x}\epsilon C \psi_{2x})
+ (\psi^T_{1 x+\hat{\mu}}\epsilon C \psi_{2x})
$$
$$
+ (\overline{\psi}_{2x}\epsilon C \overline{\psi}^T_{1x})
- (\overline{\psi}_{2 x+\hat{\mu}}\epsilon C \overline{\psi}^T_{1x})
$$
$$
\left.\left.
- (\overline{\psi}_{1x}\epsilon C \overline{\psi}^T_{2x})
+ (\overline{\psi}_{1 x+\hat{\mu}}\epsilon C \overline{\psi}^T_{2x})
\right)\right]
$$
$$
+ (\overline{\psi}_{1Rx} G_1\varphi^+_x \psi_{1Lx})
+ (\overline{\psi}_{1Lx} \varphi_x G_1  \psi_{1Rx})
$$
\be \label{eq21}
\left.
+ (\overline{\psi}_{2Rx} G_2\varphi^+_x \psi_{2Lx})
+ (\overline{\psi}_{2Lx} \varphi_x G_2  \psi_{2Rx}) \right\} \ .
\ee
 Here the summations $\sum_\mu$ always go over eight directions of the
 neighbouring sites, $C$ is the Dirac matrix for charge conjugation and
 $\epsilon = i\tau_2$ is the antisymmetric unit matrix in isospin space.
 In the scalar part of the action $m_0^2$ is the bare mass squared and
 $\lambda$ the bare quartic coupling.
 In the fermionic part $\mu_0$ is an off-diagonal Majorana mass term,
 $r$ is the Wilson parameter for removing lattice fermion doublers,
 which is usually chosen to be $r=1$, and $G_{(1,2)}$ are diagonal $2
 \otimes 2$ matrices in isospin space for the bare Yukawa-couplings:
\be \label{eq22}
G_{(1,2)} \equiv \left(
\begin{array}{cc}
G_{(1,2)u}  &  0  \\  0  &  G_{(1,2)d}
\end{array}  \right)  \ .
\ee

 The global $\rm SU(2)_L$ symmetry is acting on the fields as
$$
\varphi_x^\prime = U_L^{-1}\varphi_x \ ,
$$
$$
\psi_{(1,2)Lx}^\prime = U_L^{-1}\psi_{(1,2)Lx} \ ,
$$
\be \label{eq23}
\overline{\psi}_{(1,2)Lx}^\prime =
\overline{\psi}_{(1,2)Lx} U_L \ .
\ee
 The right-handed components of the fermion fields $\psi_{(1,2)Rx}$ and
 $\overline{\psi}_{(1,2)Rx}$ are, of course, invariant.

 The global U(1) symmetry corresponds to the conservation of the
 difference of the fermion number of $\psi_1$ minus the fermion
 number of $\psi_2$.
 (This means that if it is identified by the hypercharge $\rm U(1)_Y$,
 then $\psi_1$ and $\psi_2$ have opposite hypercharges.)

 An equivalent form of the fermionic part of the above action is
 obtained, if one introduces the mirror fermion fields by charge
 conjugation:
\be \label{eq24}
\chi_x \equiv \epsilon^{-1} C \overline{\psi}_{2x}^T \ ,  \hspace{1em}
\overline{\chi}_x \equiv \psi_{2x}^T \epsilon C  \ .
\ee
 Since under charge conjugation the left- and right-handed components
 are interchanged, and $\epsilon^{-1}U_L^T\epsilon=U_L^{-1}$, under
 $\rm SU(2)_L$ transformations we have
\be \label{eq25}
\chi_{Rx}^\prime = U_L^{-1}\chi_{Rx} \ ,  \hspace{1em}
\overline{\chi}_{Rx}^\prime = \overline{\chi}_{Rx} U_L \ ,
\ee
 and now the left-handed components $\chi_{Lx}$ and
 $\overline{\chi}_{Lx}$ are invariant.
 Omitting the index on $\psi_1$, one gets the action in terms of the
 mirror pair of fermion fields \cite{CHFER}
$$
S_{fermion} = \sum_x \left\{
  \mu_0 \left[ (\overline{\chi}_x\psi_x)
+ (\overline{\psi}_x\chi_x) \right]
\right.
$$
$$
- \half \sum_\mu \left[
  (\overline{\psi}_{x+\hat{\mu}}\gamma_\mu\psi_x)
+ (\overline{\chi}_{x+\hat{\mu}}\gamma_\mu\chi_x)
\right.
$$
$$
- r \left(
  (\overline{\chi}_x\psi_x)
- (\overline{\chi}_{x+\hat{\mu}}\psi_x)
\right.
$$
$$
\left.\left.
+ (\overline{\psi}_x\chi_x)
- (\overline{\psi}_{x+\hat{\mu}}\chi_x) \right) \right]
$$
$$
+ (\overline{\psi}_{Rx} G_\psi\varphi^\dagger_x \psi_{Lx})
+ (\overline{\psi}_{Lx} \varphi_x G_\psi  \psi_{Rx})
$$
\be \label{eq26}
\left.
+ (\overline{\chi}_{Lx} G_\chi\varphi^\dagger_x \chi_{Rx})
+ (\overline{\chi}_{Rx} \varphi_x G_\chi  \chi_{Lx}) \right\} \ .
\ee
 The Yukawa-coupling of the fermion doublet is denoted here by
 $G_\psi = G_1$, and the Yukawa-coupling of the mirror fermion
 doublet is $G_\chi = \epsilon^{-1}G_2 \epsilon$.
 This means that in $G_\chi$ the isospin components are interchanged.
 The mass term proportional to $\mu_0$ and the off-diagonal
 Wilson term multiplied by $r$ look in the second form (\ref{eq26}) not
 Majorana-like but Dirac-like.

 Note that if the doublets are degenerate, that is the Yukawa-couplings
 are proportional to the unit matrix in isospin space, then the
 $\rm SU(2)_L \otimes U(1)$ symmetry is enlarged
 to $\rm SU(2)_L \otimes SU(2)_R$ defined by
$$
\varphi_x^\prime = U_L^{-1}\varphi_x U_R \ ,
$$
$$
\psi_{(L,R)x}^\prime = U_{(L,R)}^{-1}\psi_{(L,R)x} \ ,
$$
$$
\overline{\psi}_{(L,R)x}^\prime =
\overline{\psi}_{(L,R)x} U_{(L,R)} \ ,
$$
$$
\chi_{(R,L)x}^\prime = U_{(L,R)}^{-1}\chi_{(R,L)x} \ ,
$$
\be \label{eq27}
\overline{\chi}_{(R,L)x}^\prime =
\overline{\chi}_{(R,L)x} U_{(L,R)} \ .
\ee

\subsection{ Chirality and decoupling }
 An important property of the lattice action in the previous
 subsection is the possibility of decoupling half of the fermions from
 the interacting sector \cite{ROMA1}.
 Let us formulate this in the mirror fermion langauge corresponding
 to (\ref{eq26}).
 For vanishing Yukawa-coupling of the mirror fermion doublet $G_\chi=0$
 and fermion mirror fermion mixing mass $\mu_0=0$ the action is
 invariant with respect to the Golterman-Petcher shift symmetry
\be \label{eq28}
\chi_x \rar \chi_x + \epsilon \ ,  \hspace{1em}
\overline{\chi}_x \rar \overline{\chi}_x + \overline{\epsilon} \ .
\ee
 This implies \cite{GOLPET,LINWIT} that all higher vertex functions
 containing the $\chi$-field identically vanish, and the $\chi$-$\chi$
 and $\chi$-$\psi$ components of the inverse propagator are equal to
 the corresponding components of the free inverse propagator:
\be \label{eq29}
\tilde{\Gamma}_{\psi\chi} = \mu_0 + \frac{r}{2} \hat{p}^2 \ ,
\hspace{1em}
\tilde{\Gamma}_{\chi\chi} = i\gamma \cdot \bar{p} \ ,
\ee
 where, as usual, $\bar{p}_\mu \equiv \sin p_\mu$ and
 $\hat{p}_\mu \equiv 2\sin \half p_\mu$.

 The consequence of (\ref{eq29}) is that the fermion mirror fermion
 mixing mass $\mu_0$ is not renormalized by the Yukawa-interaction
 of the $\psi$-field.
 This is very useful in numerical simulations, because the
 corresponding bare parameter (usually the fermionic hopping parameter
 $K \equiv (2\mu_0+8r)^{-1}$) is fixed, and the number of bare
 parameters to be tuned is less.

 There is also another possible interpretation of the fermion
 decoupling.
 Namely, interchanging the r\^oles of $\psi$ and $\chi$, in the case
 of $G_\psi=\mu_0=0$ the ordinary fermions are decoupled.
 This decoupling scenario is in fact a rather good approximation to the
 situation in phenomenological models with mirror fermions discussed in
 the previous section.
 This is due to the fact that all known physical fermions have very
 small Yukawa-couplings.
 The only fermion states with strong Yukawa-coupling would be the
 members of the mirror families, if they would exist.
 In fact, the smallness of the known fermion masses on the electroweak
 scale could then be explained by the approximate validity of the
 Golterman-Petcher shift symmetry.
 {\em Low energy chirality would be the consequence of the
 approximate decoupling of light fermions.}
 In this sense the mirror fermion model is natural, because the
 smallness of some of its parameters is connected to an approximate
 symmetry \cite{THOOFT}.

 Let us shortly discuss the form of the broken Golterman-Petcher
 identities.
 They are broken in general by the small Yukawa-coupling $G_\chi$, by
 small mixing mass $\mu_0$, and by the small
 $\rm SU(3) \otimes SU(2) \otimes U(1)$ gauge couplings.
 For definiteness, let us consider here the case of $G_\chi \simeq 0$.
 Consider the generating function of the connected Green's functions
 $W[\eta,\overline{\eta},\zeta,\overline{\zeta},j]$, where the
 external sources $\eta,\zeta,j$ belong, respectively, to
 $\psi,\chi,\varphi$.
 The identities obtained by shifting the $\chi$- and
 $\overline{\chi}$-fields are
$$
\overline{\zeta}_x
+ \half \sum_{\mu=1}^4 (\Delta^f_\mu + \Delta^b_\mu)
\frac{\partial W}{\partial\zeta_x}\gamma_\mu
$$
$$
+ \frac{r}{2} \sum_{\mu=1}^4 \Delta^f_\mu\Delta^b_\mu
\frac{\partial W}{\partial\eta_x}
= \mu_0 \frac{\partial W}{\partial\eta_x}
$$
$$
+ G_\chi \left(
\frac{\partial^2 W}{\partial j_{Rx}\partial\zeta_x}
+ \frac{\partial W}{\partial j_{Rx}}
\frac{\partial W}{\partial\zeta_x} \right) \Gamma^\dagger_R  \ ,
$$
$$
- \zeta_x
- \half \sum_{\mu=1}^4 \gamma_\mu (\Delta^f_\mu + \Delta^b_\mu)
\frac{\partial W}{\partial\overline{\zeta}_x}
$$
$$
+ \frac{r}{2} \sum_{\mu=1}^4 \Delta^f_\mu\Delta^b_\mu
\frac{\partial W}{\partial\overline{\eta}_x}
= \mu_0 \frac{\partial W}{\partial\overline{\eta}_x}
$$
\be \label{eq30}
+ G_\chi \left(
\frac{\partial^2 W}{\partial j_{Rx}\partial\overline{\zeta}_x}
+ \frac{\partial W}{\partial j_{Rx}}
\frac{\partial W}{\partial\overline{\zeta}_x} \right) \Gamma^\dagger_R
\ .
\ee
 Here $\Delta^f_\mu$ and $\Delta^b_\mu$ denote, respectively, forward
 and backward lattice derivatives, and real scalar field components
 $\phi_R$ $(R=0,1,2,3)$ are introduced by
$$
\half \left[ \varphi_x (1+\gamma_5)
+ \varphi^\dagger_x (1-\gamma_5) \right]
$$
\be \label{eq31}
= \phi_{0x} + i\gamma_5 \tau_r\phi_{rx} \equiv \Gamma_R\phi_{Rx} \ .
\ee

 Let us define the composite fermion field $\Psi_x$ by
\be \label{eq32}
\Gamma_R\phi_{Rx}\chi_x
= \varphi_x\chi_{Lx} + \varphi^\dagger_x\chi_{Rx} \equiv \Psi_x \ .
\ee
 The mixed $\psi$-$\Psi$ two point function is
\be \label{eq33}
\langle \psi_y \Psi_x \rangle \equiv \frac{1}{\Omega}
\sum_k e^{ik \cdot (y-x)} \tilde{\Delta}^{\psi\Psi}_k \ .
\ee
 Taking derivatives at zero sources, with the notation
 $\langle \phi_{0x} \rangle \equiv v$ for the vacuum expectation
 value one obtains, for instance,
$$
0 = (\mu_0 + \frac{r}{2} \hat{k}^2) \tilde{\Delta}^{\psi\psi}_k
+ \tilde{\Delta}^{\psi\chi}_k (i\gamma \cdot \bar{k} + G_\chi v)
+ G_\chi \tilde{\Delta}^{\psi\Psi}_k \ ,
$$
$$
1 = (\mu_0 + \frac{r}{2} \hat{k}^2) \tilde{\Delta}^{\chi\psi}_k
$$
\be \label{eq34}
+ \tilde{\Delta}^{\chi\chi}_k (i\gamma \cdot \bar{k} + G_\chi v)
+ G_\chi \tilde{\Delta}^{\chi\Psi}_k \ .
\ee
 For $G_\chi=0$ this is equivalent to (\ref{eq29}).
 The case of small gauge couplings can be treated similarly.

\subsection{ A simple $SU(2)_L \otimes SU(2)_R$ model }
 The lattice actions of Yukawa-models in the form (\ref{eq21}) or
 (\ref{eq26}) can be used for numerical simulation studies of chiral
 Yukawa-models.
 In order to apply Monte Carlo simulation methods one needs, however,
 a fermion determinant which is positive.
 For instance, in the Hybrid Monte Carlo algorithm \cite{DKPR} one
 has to duplicate the number of fermionic degrees of freedom.
 Let us denote the fermion matrix corresponding to (\ref{eq26}) by
 $Q$, then the replica fermions have $Q^\dagger$, and the fermion
 determinant is $\det(QQ^\dagger)$, which is positive.
 Due to the adjoint, for the replica fermions $\psi_x$ describes
 a mirror fermion doublet and $\chi_x$ an ordinary fermion doublet.
 By charge conjugation as in (\ref{eq24}) one can transform the
 $\psi$-field of replica fermions to an ordinary doublet, and
 the $\chi$-field of replica fermions to a mirror doublet.
 In this way one can consider two doublets described by the
 $\psi$-fields and two mirror doublets described by the $\chi$-fields.
 For simplicity, let us consider only degenerate doublets, that is,
 let the Yukawa-couplings $G_{(\psi,\chi)}$ be proportional to the
 unit matrix in isospin space.
 In this case the Hybrid Monte Carlo simulation describes two
 equal mass degenerate doublets plus two equal mass degenerate mirror
 doublets with exact global $\rm SU(2)_L \otimes SU(2)_R$ symmetry.
 In the phase with spontaneously broken symmetry the mass of the
 doublets is proportional to $G_\psi$, and the mass of the mirror
 doublets to $G_\chi$.

 As it was noted in the previous subsection, the limit $G_\chi=\mu_0=0$
 is particularly interesting, because it has less bare parameters to
 tune.
 In this case the $\chi$-fields are exactly decoupled, and one is
 left in Hybrid Monte Carlo with two degenerate fermion doublets
 described by the $\psi$-fields.
 This is the simplest model of heavy fermion doublets one can simulate
 by present day fermionic simulation techniques \cite{SU2XSU2}.
 Besides the two bare parameters of the pure scalar sector
 $(m_0^2,\lambda)$ there is only one additional bare parameter
 ($G_\psi$) in the fermionic part of the action:
$$
S_{fermion} = \sum_x \left\{ -\half \sum_\mu \left[
  (\overline{\psi}_{x+\hat{\mu}}\gamma_\mu\psi_x)
\right.\right.
$$
$$
+ (\overline{\chi}_{x+\hat{\mu}}\gamma_\mu\chi_x)
$$
$$
\left. - \left(
  (\overline{\chi}_x\psi_x)
- (\overline{\chi}_{x+\hat{\mu}}\psi_x)
+ (\overline{\psi}_x\chi_x)
- (\overline{\psi}_{x+\hat{\mu}}\chi_x) \right) \right]
$$
\be \label{eq35}
\left. + G_\psi \left[
  (\overline{\psi}_{Rx} \varphi^+_x \psi_{Lx})
+ (\overline{\psi}_{Lx} \varphi_x \psi_{Rx}) \right] \right\} \ .
\ee
 To have the smallest possible number of parameters is very important,
 in order to keep parameter tuning as easy as possible.

 By a further duplication of the fermion fields one can also
 simulate four degenerate fermion doublets, which correspond to
 a heavy degenerate family.
 The $\rm SU(4)_{Pati-Salam}$ symmetry \cite{PATSAL} of the four
 $\psi$-doublets, including $\rm SU(3)_{colour}$ for the quarks,
 is exact in the continuum limit, but for finite lattice spacings
 it is broken by the off-diagonal Wilson-terms which mix the $\psi$- and
 $\chi$-fields.
 Note, however, that there is an exact SU(4) symmetry also at finite
 lattice spacings, if one transforms the $\psi$- and $\chi$-fields
 simultaneously.
 Of course, the $\chi$'s are mirror fermion fields, which mix with the
 $\psi$'s through the nonzero $\tilde{\Gamma}_{\psi\chi}$ in
 (\ref{eq29}).
 This mixing goes to zero only in the continuum limit.
 The decoupling in the continuum limit is exact in the Yukawa-model,
 but for nonzero gauge couplings decoupling the $\chi$'s in a gauge
 invariant way does not work.

 Another way of simulating a heavy degenerate fermion family with
 only two pairs of $(\psi,\chi)$-doublet fields is to choose
 $G_\chi = \pm G_\psi$ (the opposite sign is preferred by the study
 of the $K=0$ limit \cite{LIMAMO}).
 Since without $\rm SU(3) \otimes U(1)$ gauge fields the mirror
 fermion doublets are equivalent to ordinary fermion doublets, this
 describes the same model in the continuum limit as the one with
 twice as much fields and decoupling.
 In this case, however, the fermion hopping parameter $K$ has to be
 tuned, too, which can be worse than having more field components per
 lattice sites.
\begin{figure}[tb]
\vspace{9cm}
\caption{ \label{fig1}
 Phase structure of the $\rm SU(2)_L\otimes SU(2)_R$ symmetric Yukawa
 model at $\lambda=\infty$ and $G_\chi=\mu_0=0$ in the
 ($G_\psi,\,\kappa$)-plane.
 Open circles denote points in the PM~phase, crosses represent
 points in the FM~phase.
 The points in the AFM and FI~phases are denoted by full circles and
 open squares, respectively.
 The dashed lines labelled~R,S,T each show the range of~$\kappa$ used
 for a systematic scan of renormalized parameters at fixed~$G_\psi$.
 The crosses along those lines denote the kappa values where the
 minimum scalar mass in the broken phase is encountered.
 Solid lines connect the critical values for $\kappa$ estimated from
 the behaviour of the magnetization on $4^3\cdot8$.
 Dashed lines around the FI~phase show the expected continuation of the
 critical lines. }
\end{figure}

\subsection{ Phase structure }
 The first step in a recent numerical simulation of the
 $\rm SU(2)_L \otimes SU(2)_R$ symmetric Yukawa-model with $N_f=2$
 fermion doublets in the decoupling limit $G_\chi=0$ \cite{SU2XSU2}
 was to check the phase structure at infinite bare quartic coupling
 $\lambda=\infty$.
 On the basis of experience in several different lattice Yukawa models
 \cite{GOLREV,SHIREV,LIMOWI}, this is expected to possess several phase
 transitions between the {\it ``ferromagnetic'' (FM)}, {\it
 ``antiferromagnetic'' (AFM)}, {\it ``paramagnetic'' (PM)} and {\it
 ``ferrimagnetic'' (FI)} phases.
 The resulting picture in the ($G_\psi,\kappa$)-plane is shown in
 fig.~\ref{fig1}.
 ($\kappa \equiv (1-2\lambda)/(m_0^2+8)$ is the bare parameter which
 is usually taken in numerical simulations instead of $m_0^2$.)
\begin{figure}[tb]
\vspace{9cm}
\caption{ \label{fig2}
 The cut-off dependent allowed region in the ($G_{R\psi}^2,g_R$) plane
 for cut-off values equal to some multiples of the Higgs-boson mass
 $m_\phi \equiv m_{R\sigma}$.
 In the Yukawa-model describing two degenerate heavy fermion doublets
 without gauge couplings the perturbative 1-loop $\beta$-functions are
 assumed. }
\end{figure}

\subsection{ Allowed region in renormalized couplings }
 An important question for numerical simulations is the determination
 of the nonperturbative cut-off dependent {\em allowed region} in the
 space of renormalized quartic and Yukawa-couplings.
 If the continuum limit of Yukawa-models is trivial, then there
 are cut-off dependent upper bounds on both the renormalized quartic
 and Yukawa-couplings, which tend to zero in the continuum limit.
 In pure $\phi^4$ models the upper bound is qualitatively well
 described by the 1-loop perturbative $\beta$-function, if the
 Landau-pole in the renormalization group equations is assumed to
 occur at the scale of the cut-off.
 The same might be true for scalar-fermion models with Yukawa-couplings.
 For instance, in the model with $\rm SU(2)_L \otimes SU(2)_R$ symmetry
 and $N_f$ degenerate fermion doublets the 1-loop $\beta$-functions for
 the quartic ($g_R$) and Yukawa- ($G_{R\psi}$) couplings are:
$$
\beta_{g_R} = \frac{1}{16\pi^2} \left(
4g_R^2 + 16N_fg_RG_{R\psi}^2 - 96N_fG_{R\psi}^4 \right) \ ,
$$
\be \label{eq36}
\beta_{G_{R\psi}} = \frac{1}{16\pi^2} \cdot
4N_fG_{R\psi}^3 \ .
\ee

 Since in the region where $G_{R\psi}^2 \gg g_R$ the 1-loop
 $\beta$-function of the quartic coupling $\beta_{g_R}$ is negative,
 besides the upper bounds there is also a lower bound on $g_R$ for fixed
 $G_{R\psi}$, which is called in the literature {\it vacuum stability
 bound} \cite{SHER}.
 On the lattice, if one assumes the qualitative behaviour of the 1-loop
 $\beta$-function to be valid also nonperturbatively, the vacuum
 stability lower bound occurs at zero bare quartic coupling $\lambda=0$,
 whereas the upper bound at $\lambda=\infty$ \cite{LMMW,LMMWTA,SHENTA}.
 Negative $\lambda$-values are excluded, because there the path
 integral is divergent.
 For the 1-loop $\beta$-functions in (\ref{eq36}) with $N_f=2$
 the bounds in the plane of ($G_{R\psi}^2,g_R$) are shown in
 fig.~\ref{fig2}.
\begin{figure}[tb]
\vspace{9cm}
\caption{ \label{fig3}
 The fermion mass in lattice units $\mu_{R\psi}$ plotted versus
 $\kappa$ for $G_\psi=0.3$ (triangles) and $G_\psi=0,6$ (squares) on
 lattices of size $4^3\cdot8$ (open symbols) and $6^3\cdot12$
 (filled-in symbols).
 Errorbars are omitted when the variation is of the size of the symbols.
 It is seen that larger bare couplings $G_\psi$ in general yield larger
 fermion masses. }
\end{figure}

 These curves have to be confronted with the results of the numerical
 simulations.
 It turned out (see fig.~\ref{fig3}) that on $4^3 \cdot 8$ and
 $6^3 \cdot 12$ lattices the fermion mass in lattice units tends to
 zero, if one is approaching the FM-PM phase transition from above
 (i.~e. from the FM-side).
 This is as expected in the continuum limit on an infinite lattice.
\begin{figure}[tb]
\vspace{9cm}
\caption{ \label{fig4}
 The scalar mass in lattice units $m_{R\sigma}$ plotted versus $\kappa$.
 The explanation of symbols is similar to the previous figure.
 When approaching the phase transition the scalar masses increase again
 after going through a minimum. }
\end{figure}
\begin{figure}[tb]
\vspace{9cm}
\caption{ \label{fig5}
 The obtained numerical estimate of the upper bound on the renormalized
 quartic coupling $g_R$ (or Higgs-boson mass $m_{R\sigma}$) as a
 function of the renormalized Yukawa-coupling squared $G_{R\psi}^2$.
 Two points with the smaller errors are on $6^3 \cdot 12$, the
 third one on $8^3 \cdot 16$ lattice.
 The lines are the upper bounds according to 1-loop perturbation
 theory at a cut-off $\Lambda \simeq 3m_{R\sigma}$ and $4m_{R\sigma}$,
 respectively. }
\end{figure}

 The behaviour of the Higgs-boson mass $m_{R\sigma}$ on finite lattices
 is more involved.
 This is because the decrease of $m_{R\sigma}$ is stopped by a minimum,
 and instead of a further decrease there is an increase (see
 fig.~\ref{fig4}).
 This can be understood as a finite size effect.
 The value at the minimum is smaller on larger lattices.
 Nevertheless, for increasing bare Yukawa-coupling $G_\psi$ the
 required lattice size is growing.
 Reasonably small masses $m_{R\sigma} \simeq 0.5-0.7$ can be
 achieved at $G_\psi=0.3$ on $6^3 \cdot 12$, at $G_\psi=0.6$ on
 $8^3 \cdot 16$ lattices \cite{SU2XSU2}.
 At $G_\psi=1.0$ presumably lattices with spatial extension of
 at least $16^3$ are necessary.
 Taking the values of the Higgs-boson mass at $\kappa$-values above the
 minimum, one obtains for the upper limit on the renormalized quartic
 coupling the estimates in fig.~\ref{fig5}.
 These agree within errors with the 1-loop estimates, although the
 values of the renormalized couplings are close to the tree
 unitarity limits.
 The continuation of the upper bound on $g_R$ towards larger
 Yukawa-couplings can only be obtained on larger lattices.
 Particularly interesting is the behaviour in the vicinity of
 the FM-FI phase transition near $G_\psi=1.0-1.5$.
 A further interesting question is the phase structure near
 $\lambda \simeq 0$, which has an influence on the vacuum stability
 lower bound.

\begin{acknowledge}
 It is a pleasure to thank the organizers for giving me the opportunity
 to participate in this workshop.
 I benefited a lot from the discussions with the participants in a
 very pleasant atmosphere.
\end{acknowledge}

%%%%%%%%%%%%%%%%%%%%%%%%%%%%%%%%%%%%%%%%%%%%%%%%%%%%%%%%%%%%%%%%%%%%%%%%

\end{document}